\documentclass[12pt,a4paper]{article}
\RequirePackage{lineno}
\usepackage{epsfig}
\usepackage{graphicx}                  
\usepackage{siunitx}
\usepackage[utf8]{inputenc}
\usepackage{ae}
\usepackage{amsmath}
\usepackage{amssymb}
\usepackage{graphics}
\usepackage{dcolumn}
\usepackage{bm}
\usepackage{hyperref}
\usepackage[symbol]{footmisc}
\usepackage{setspace}
\usepackage{ragged2e}
\usepackage{wrapfig}
\DeclareUnicodeCharacter{2212}{-}
\begin{document}

\centering

{\Large
			{\bf Charged particle to photon ratio in heavy-ion collisions at RHIC: a different point of view}}
\date{}


\centering
{Mohammad Asif Bhat\footnote[1]{Corresponding Author:
asifqadir@jcbose.ac.in; bhatasif305@gmail.com} and Supriya Das}

\vspace*{0.5cm}
{Department of Physics and Centre for Astroparticle Physics and Space Science (CAPSS), ~Bose Institute, EN-80, Sector V, Kolkata 700091, India}

\vspace*{0.5cm}
\centering{\bf Abstract}
\justify

We report an alternative explanation of the variation of charged particle to photon ratio in Cu+Cu collisions and Au+Au collisions from 1.4 $\pm$ 0.1 to 1.2 $\pm$ 0.1 at $\sqrt{\it{s}_{\rm NN}}$ = 62.4 GeV and 200 GeV respectively in STAR experiment at RHIC. Based on the theoretical predictions and experimental results we argue that the additional contribution of direct photons at $\sqrt{\it{s}_{\rm NN}}$ = 200 GeV in both Cu+Cu and Au+Au collisions are improving the ratio $\frac{N_{ch}}{N_{\gamma}}$ to 1.2 $\pm$ 0.1 from 1.4 $\pm$ 0.1. 

\section{Introduction}
\label{1}
High energy experiments are being performed for last so many years to study the Quark Gluon Plasma (QGP) which is a strongly interacting hot and dense medium of quarks and gluons having very short life time. Measurement of the multiplicity of charged particles provides the information about the underlying physics processes occuring during the medium formation such as fragmentation~\cite{Ke:2020clc,Goncalves:2019uod} and hadronization~\cite{He:2019vgs,Cao:2015hia}. Measurement of photon multiplicity provides the complementary information to that of charged particles thus helps in understanding of QGP medium~\cite{Casalderrey-Solana:2019ubu,PHENIX:2018lia}.
\par
The measurement of the particle density in pseudorapidity $\eta$ is the conventional way of describing particle production in heavy ion collisions. The information of energy density, initial temperature, and velocity of sound in the medium formed in the collisions is provided by the particle density in pseudorapidity within the framework of certain model assumptions~\cite{Bjorken:1982qr}. The longitudinal flow and rescattering effects~\cite{Klay:2001tf} strongly modifies the widths of the pseudorapidity distributions. The relative importance of soft versus hard processes in particle production can be studied by studying the variation of particle density in $\eta$ with collision centrality, expressed in terms of the number of participating nucleons $N_{part}$ and/or the number of binary collisions $N_{coll}$. The test ground for various particle production models, such as those based on ideas of parton saturation~\cite{Gribov:1984tu} and semiclassical QCD, also known as the color glass condensate (CGC)~\cite{McLerran:1993ni} is provided by the particle density in pseudorapidity $\eta$. The particle production mechanism could be different in different regions of pseudorapidity at Relativistic Heavy-ion Collider (RHIC). A significant increase in charged particle production normalized to the number of participating nucleons has been observed from peripheral to central Au+Au collisions~\cite{Abbott:2003ba} at midrapidity. This is because of the onset of hard scattering processes, which scales with the number of binary collisions. However, the total charged particle multiplicity per participant pair, integrated over the whole pseudorapidity range, is independent of centrality in Au+Au collisions. The centrality dependence of particle production at midrapidity reflects the increase of gluon density due to the decrease in the effective strong coupling constant in the framework of the color glass condensate picture of particle production. 
\par 
The detailed study of the increase in particle production at midrapidity with increase in the center-of-mass energy has been done at Relativistic Heavy-ion Collider (RHIC)~\cite{Back:2001bq}. The multiplicity of hadrons and their energy, centrality and rapidity dependence from experimental data so far have been consistent with the approach based on ideas of parton saturation. It has been argued that this onset of saturation occurs somewhere in the center-of-mass energy $\sqrt{\it{s}_{\rm NN}}$ = 17 GeV to 130 GeV~\cite{Kharzeev:2001yq}.\par
\section{Photon and charged particle production at forward rapidities}
\label{2}
The number of particles (both charged particles and photons) produced per participant pair as a function of $\eta-y_{beam}$, where $y_{beam}$ is the beam rapidity, is observed
to be independent of beam energy at forward rapidities. This phenomenon is known as limiting fragmentation~\cite{Beckmann:1981gc}, which is the longitudinal scaling of particle production at forward rapidties. The centrality independent limiting fragmentation behavior is observed to be followed by the inclusive photon production (primarily from decay of $\pi^{0}$ ) at $\sqrt{\it{s}_{\rm NN}}$ = 62.4 GeV. The centrality dependent behavior of limiting fragmentation is also observed to be followed by the inclusive charged particles at 19.6 GeV and 200 GeV. 
The centrality dependent limiting fragmentation for inclusive charged particles may be due to the baryons coming from nuclear
remnants and baryon transport~\cite{Adams:2005cy}, both of which change with centrality. 
\par 
The ratios of the number of charged particles to photons in the pseudorapidity range $-3.7 < \eta < -2.3$ are found to be $1.4 \pm 0.1$ and $1.2 \pm 0.1$ for $\sqrt{\it{s}_{\rm NN}}$ = 62.4 GeV and 200 GeV, respectively in Cu+Cu and Au+Au collisions~\cite{Abelev:2009cy}. Theoretically the ratio should be one in any system at any energy if we consider only decay photons as majority of the particles produced in any collision are pions and neutral pions decay into two photons. The ratio here is varying as a function of collision energy but not with the system size. It has been conjuctured that baryons contribute to charged particles~\cite{Bearden:2003hx,Kharzeev:1996sq}, while mesons are the dominant contributors to
photon production in a given system at given energy so the ratio is more than one. Also the contribution of baryons to charged particles is suppressed as we go to the higher energies so the ratio is approaching to unity.
\par In this paper we will try to convince that the energy dependence of the charged particle to photon ratio may be due to the varying contribution of the direct photons keeping the baryon contribution fixed based on the theoretical model predictions and experimental results.\par
The paper is organized in the following way:\\
Section~\ref{1} is the introduction of the paper and section~\ref{2} discusses the photon and charged particle production at forward rapidities. Section~\ref{3} contains the results discussing the evidences supporting our argument that the energy dependence of the charged particle to photon ratio may be due to the varying contribution of the direct photons keeping the baryon contribution fixed. In section~\ref{4} we have summarized the content of the paper.
\section{Results supporting our argument}
\label{3}
Decay photons are those photons which are produced by the decay of hadrons while direct photons are those which are produced not by the decay of hadrons. Direct photons are divided into two catogories:\\(1) Prompt direct photons: prompt direct photons are produced by the initial hard scattering of partons in a  collision such as Compton scattering of partons~\cite{Owens:1986mp}.\\(2) Thermal direct photons: thermal direct photons are emitted by the hot Quark Gluon Plasma in the form of radiation or by quark-antiquark annihilation~\cite{Linnyk:2013hta}.\par
The results supporting our argument that the energy dependence of the charged particle to photon ratio may be due to the varying contribution of the direct photons keeping the baryon contribution fixed are the following:
\subsection{Increase in the direct photon cross section with the center of mass energy} It has been observed that the direct photon cross section increases with increase in the center of mass energy. The experimental result which is available to us is direct photon production in S+Au collisions at $\sqrt{\it{s}_{\rm NN}}$ = 19 GeV at WA80 experiment taking S+Au same as Au+Au and Au+Au collisions at $\sqrt{\it{s}_{\rm NN}}$ = 200 GeV measured by PHENIX experiment at RHIC.
\begin{figure}[htb]
	\includegraphics[width=0.9\textwidth]{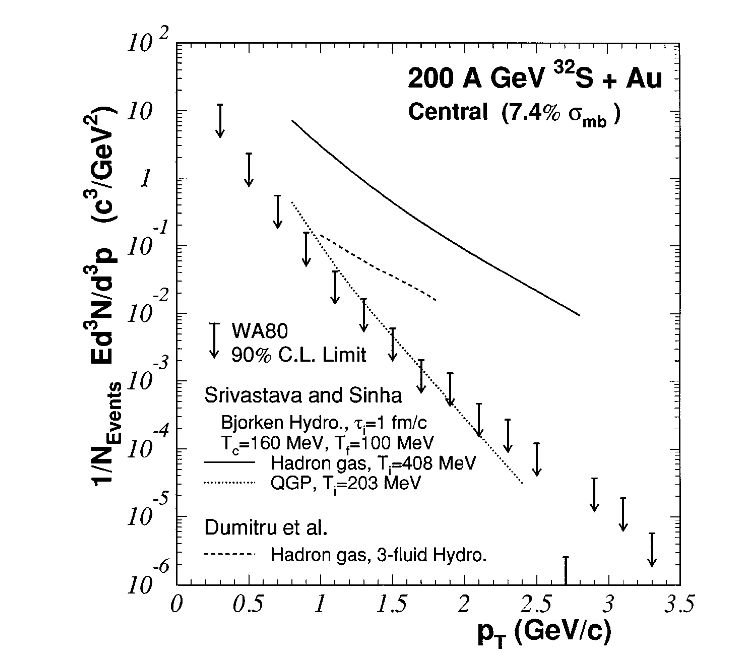}
	\caption{\small Upper limits on direct photon production in central S+Au collisions at $\sqrt{\it{s}_{\rm NN}}$ = 19 GeV as measured by the WA80 experiment~\cite{Albrecht:1995fs}. The bars at the foot of each arrow mark the 90\% C.L. upper limit on the differential yield per event (the lengths of the arrows have no significance).}
	\label{1}
\end{figure}
\begin{figure}
	\includegraphics[width=0.9\textwidth]{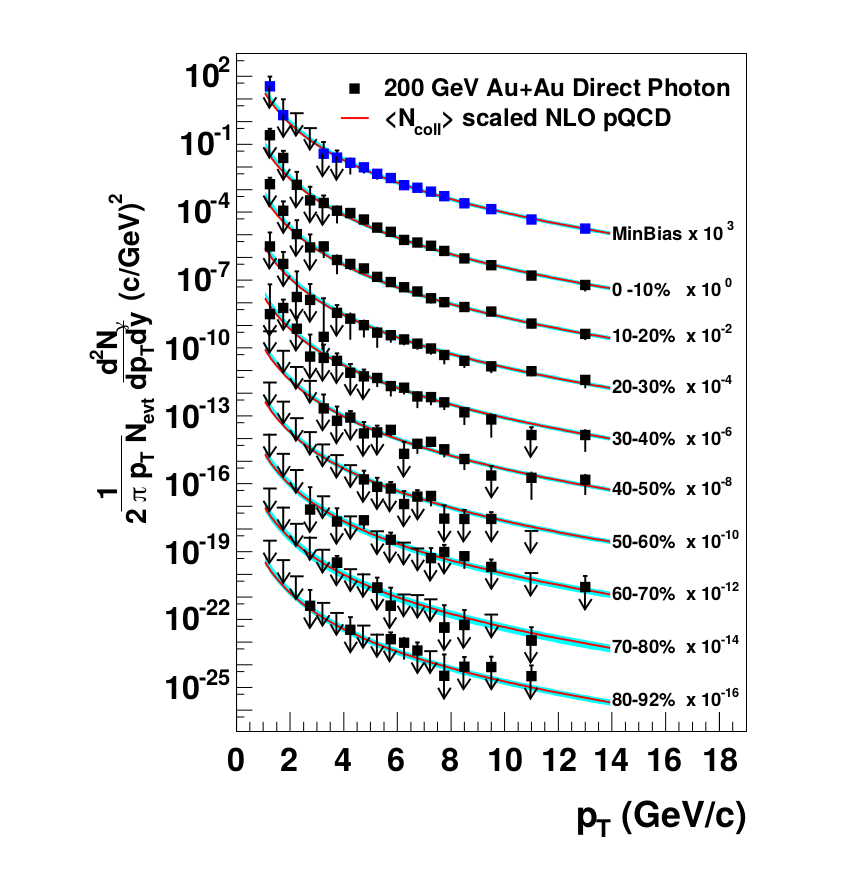}
	\caption{\small Direct photon invariant yields as a function of transverse momentum for different centrality selections and minimum bias Au+Au collisions at $\sqrt{\it{s}_{\rm NN}}$ = 200 GeV~\cite{Adler:2005ig}. The vertical error bar on each point indicates the total error. Arrows indicate measurements consistent with zero yield with the tail of the arrow indicating the 90\% confidence level upper limit. The solid curves are pQCD predictions described in the text.}
	\label{2}
\end{figure}
\begin{figure}
	\includegraphics[width=0.9\textwidth]{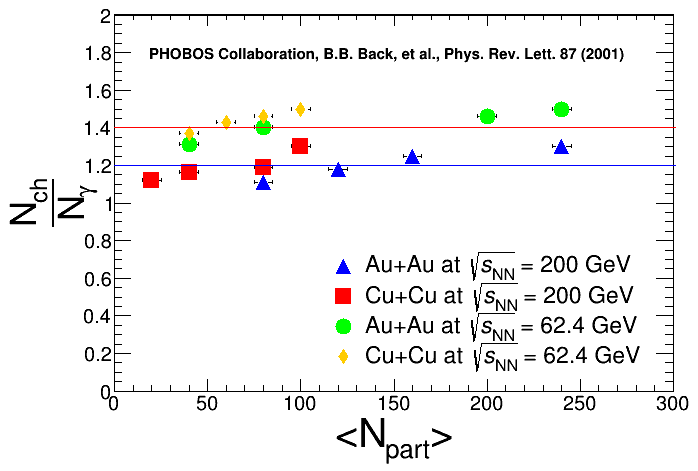}
	\caption{\small The ratio of number of charged particles to the number of  photons as a function of average number of participating nucleons $<N_{part}>$ for Au+Au and Cu+Cu collisions at $\sqrt{\it{s}_{\rm NN}}$ = 200 GeV and 62.4 GeV for $−3.7 < \eta < −2.3$.}
	\label{3}
\end{figure}
The comparison between the two results in Fig~\ref{1} and Fig~\ref{2} shows that both multiplicity and energy of the produced direct photons increase with the center of mass energy.\par 	
The ratios of the number of charged particles to photons in Fig~\ref{3} varying from 1.4 to 1.2 while going from $\sqrt{\it{s}_{\rm NN}}$ = 62.4 GeV to 200 GeV in both Cu+Cu and Au+Au collisions may be due to the reason that direct photons at $\sqrt{\it{s}_{\rm NN}}$ = 62.4 GeV in both Cu+Cu and Au+Au collisions are produced in small quantity and with less energy so that they are unable to reach the forward region $(-3.7 < \eta < -2.3)$ where the Photon Multiplicity Detector is placed, therefore only decay photons are contributing to the photon multiplicity. On the other hand direct photons produced at $\sqrt{\it{s}_{\rm NN}}$ = 200 GeV in both Cu+Cu and Au+Au collisions are produced in large quantity and with high energy so that they are able to reach Photon Multiplicity Detector and are contributing to the photon multiplicity along with decay photons. This additional contribution of direct photons at $\sqrt{\it{s}_{\rm NN}}$ = 200 GeV in both Cu+Cu and Au+Au collisions may be the reason of improving the ratio $\frac{N_{ch}}{N_{\gamma}}$ to 1.2 $\pm$ 0.1 from 1.4 $\pm$ 0.1 keeping the energy to charged particle production fraction same.
\subsection{Existence and enhancement of direct photons in the forward region}
\begin{figure}[htb]
	\includegraphics[width=1\textwidth]{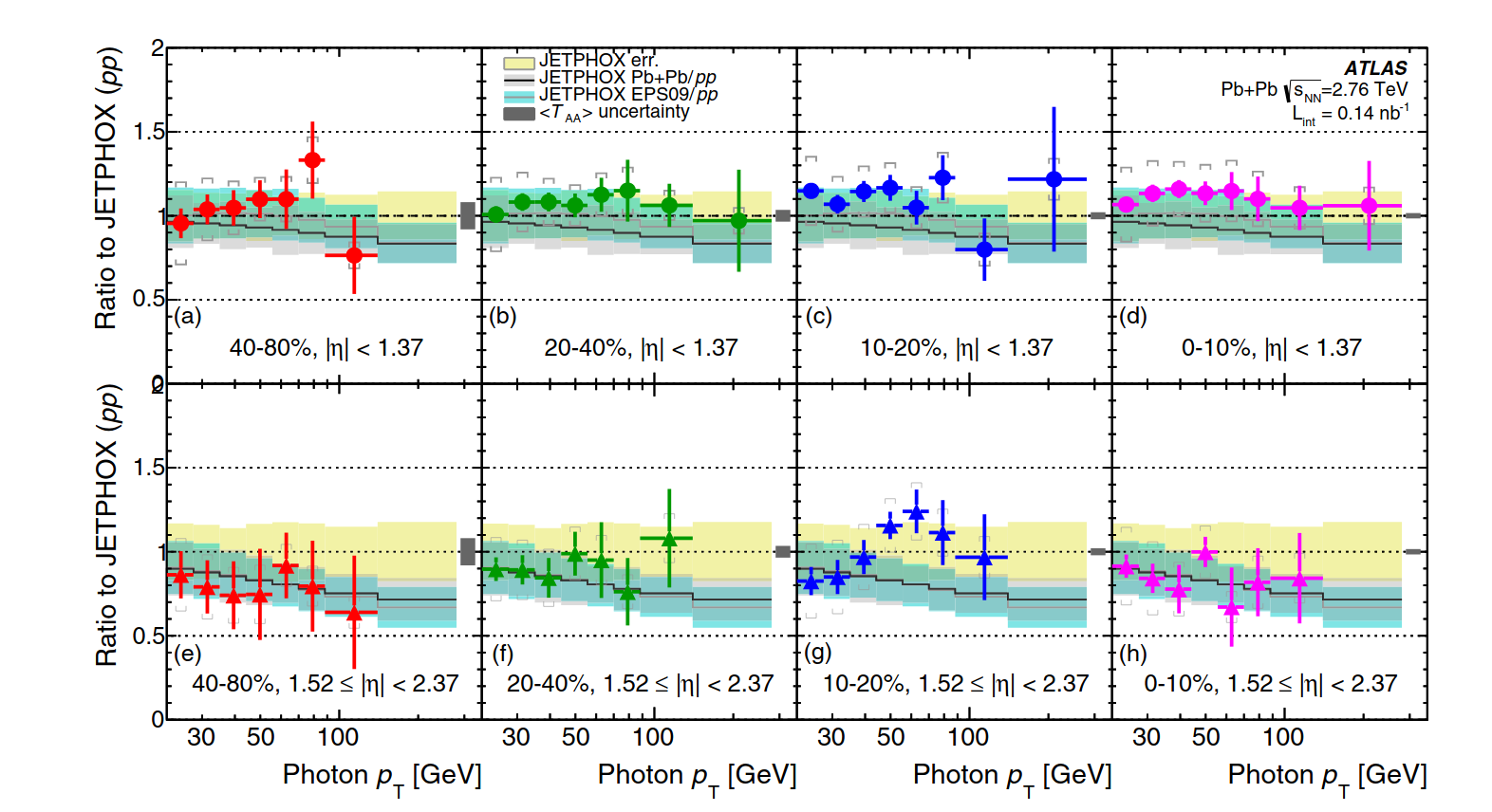}
	\caption{\small The normalized yields of prompt photons in Pb+Pb collisions at $\sqrt{\it{s}_{\rm NN}}$ = 2.76 TeV divided by JETPHOX predictions for p+p collisions as a function of $p_{T}$ in central (top panels, $|\eta| < 1.37$) and forward (bottom panels, $1.52 < |\eta| < 2.37$) rapidities in different centrality selections from peripheral (left panels) to central (right panels). The combined scale and PDF uncertainty on the JETPHOX calculation is shown by the yellow area. Two other scenarios of JETPHOX calculation divided to the same JETPHOX p+p is shown, including the isospin effect from Pb+Pb collisions (black curve), and nuclear modification of the PDFs (blue area)~\cite{Aad:2015lcb}.}
	\label{6}
\end{figure}

Direct photons are radiated by the electric current of the projectile
quarks, which mostly stay in the fragmentation region of the beam, and tend to form a peak at forward rapidities as shown in Fig~\ref{4}. But at very large $p_T$ and $\eta$ the kinematic limit pushes photon radiation to more central rapidities
and the peak at forward rapidities will be replaced by a kind of plateau at central rapidities. Also gluons are radiated via nonabelian mechanisms by the color current across the whole rapidity interval and tend
to form a plateau at midrapidity. The result shown below in Fig~\ref{5} is obtained from the Golec-Biernat and W$\ddot u$sthoff (GBW) model which is a popular parametrization for the q$\bar{q}$ dipole cross-section on a nucleon target~\cite{JalilianMarian:2007bg,Jeon:2002mf}. The Fig~\ref{5}
\begin{figure}\vspace{-4.7cm}
	\includegraphics[width=1\textwidth]{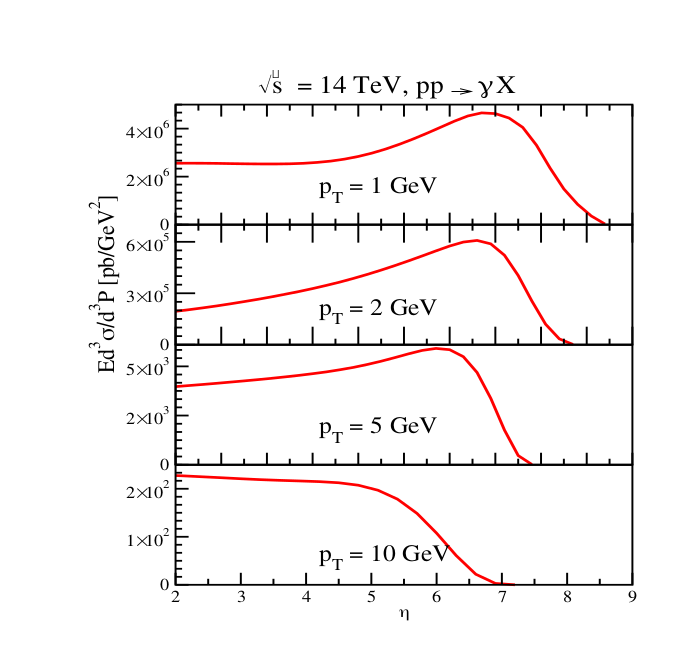}
	\caption{\small Invariant cross-section for direct photon production in pp collisions at LHC as a function of rapidity $\eta$ calculated with the Golec-Biernat and W$\ddot u$sthoff (GBW) color dipole model for various fixed $p_{T}$ ~\cite{Rezaeian:2009it}.}
	\label{4}
\end{figure}
\begin{figure}
	\includegraphics[width=1\textwidth]{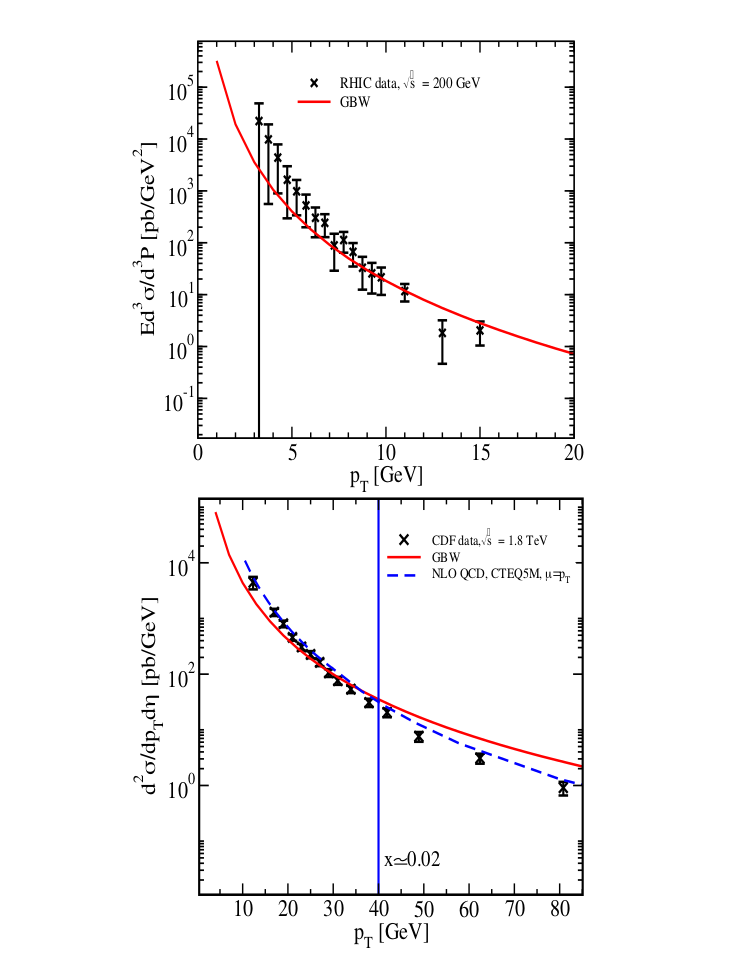}
	\caption{\small Direct photon spectra obtained from the GBW dipole
		model at the RHIC and CDF energies for pp collisions~\cite{Rezaeian:2009it}. Experimental data are from the PHENIX experiment~\cite{Adler:2006yt} at $\eta$ = 0,
		and from the CDF experiment~\cite{Acosta:2004bg,Abe:1994rra} at $\eta < 0.9$. The error bars are the linear sum of the statistical and systematic uncertainties.}
	\label{5}
\end{figure}
shows the direct photon yield predicted by the (GBW) color dipole model is almost in good agreement with the RHIC data in pp collisions at $\sqrt{\it{s}}$ = 200 GeV.
The experimental result shown in Fig~\ref{6} is from ATLAS experiment showing the enhancement of direct photon yield in the more forward region $1.52 \le |\eta| < 2.37$ in the centrality class 10-20\% in Pb+Pb collisions at $\sqrt{\it{s}_{\rm NN}}$ = 2.76 TeV.
The theoretical prediction and experimental results showing that direct photons not only exist in the forward region but are enhanced in the more forward region supports our argument that the additional contribution of direct photons at $\sqrt{\it{s}_{\rm NN}}$ = 200 GeV in both Cu+Cu and Au+Au collisions.
\subsection{Estimation of direct photon contribution}
It was observed in WA80 experiment that for O+Au collisions direct photons constituted no more than 15\% of the inclusive photon yield at each $p_{T}$ in the range $0.4 < p_{T} < 2.4$ GeV$/c$.
\begin{figure}\vspace{-2cm}
	\centering
	\includegraphics[width=1\textwidth]{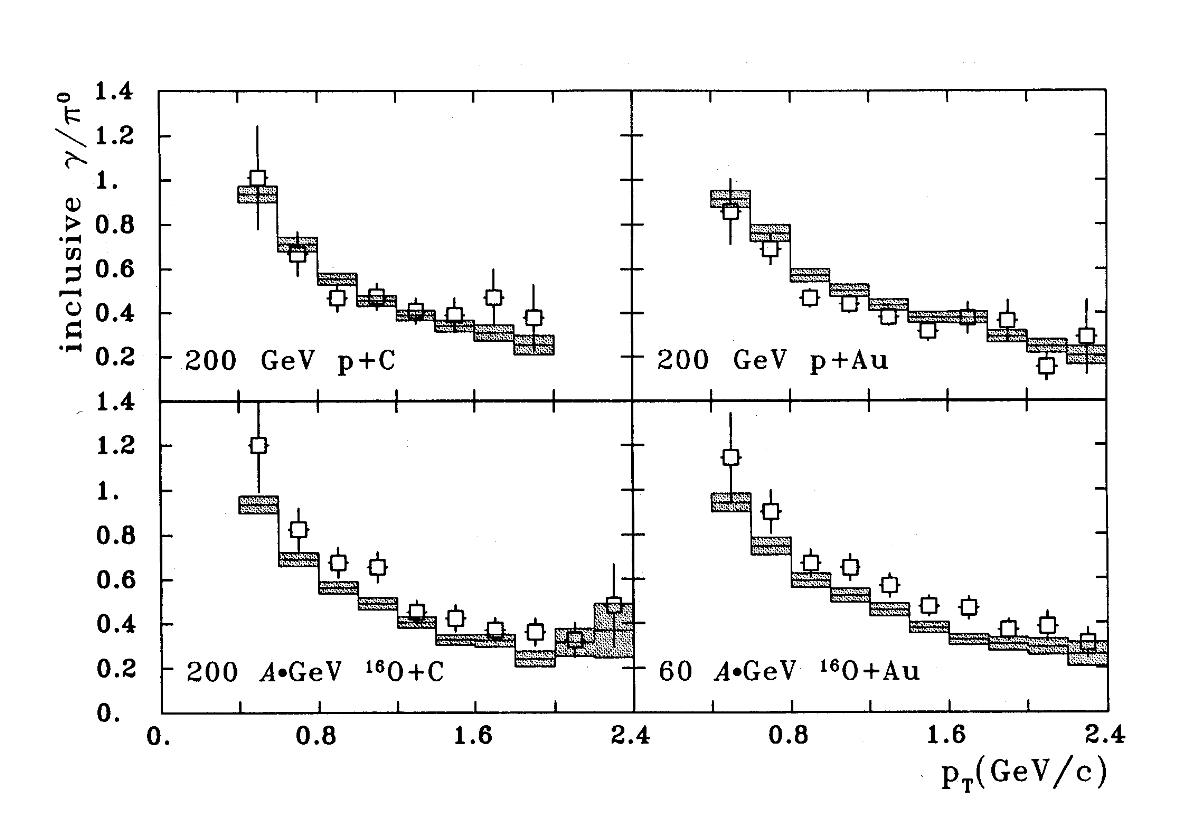}
	\caption{\small Ratio of inclusive photon cross section to $\pi^{0}$ cross section as a function of transverse momentum for 200 GeV p+C, 200 GeV p+Au , 200 A$\cdot$GeV $O^{16}$+C, and 60 A$\cdot$GeV $O^{16}$+Au events under minimum-bias trigger conditions. The squares indicate the data, and the shaded histograms are the Monte Carlo results representing hadronic decays.~\cite{Albrecht:1990jq}}
	\label{7}
\end{figure}
Fig~\ref{7} shows the collision systems 200 GeV p+C, 200 GeV p+Au, 200 A$\cdot$GeV $O^{16}$+C, and 60 A$\cdot$GeV $O^{16}$+Au under minimum-bias trigger conditions. Experimental data in the Fig~\ref{7} is indicated by the squares and Monte Carlo results are indicated by the shaded histograms based on measured $\pi^{0}$ spectra and represent the hadronic background. The direct photon contribution can be attributed to the difference between data and the histograms. The $p_{T}$ dependence of the data is reproduced by the Monte Carlo simulations for all four systems. The data for the $O^{16}$ induced reactions are approximately 10\% above the prediction, but this finding is still within the systematic uncertainty~\cite{Akesson:1989tm} of the data. After taking all the systematic uncertainties into consideration the contribution of direct photons is approximately about 15\% of the neutral meson production~\cite{Albrecht:1990jq}.\par
Based on this observation we assume that at $\sqrt{\it{s}_{\rm NN}}$ = 62.4 GeV the direct photons do not reach the forward rapidities where Photon Multiplicity Detector is situated, thus keeping the ratio of $\frac{N_{ch}}{N_{\gamma}}$ high. But at $\sqrt{\it{s}_{\rm NN}}$ = 200 GeV if 15\% additional contribution from direct photons is added to the yield of decay photons, the experimentally observed ratio $\frac{N_{ch}}{N_{\gamma}}$ = 1.2 for both Cu+Cu and Au+Au could be explained.
\section{Summary}
\label{4}
In this work we report an alternative explanation of the variation of charged particle to photon ratio in Cu-Cu collisions and Au-Au collisions from 1.4 $\pm$ 0.1 to 1.2 $\pm$ 0.1 at $\sqrt{\it{s}_{\rm NN}}$ = 62.4 GeV and 200 GeV respectively in STAR experiment at RHIC. The experimental results such as increase in the direct photon cross section with the center of mass energy, existence and enhancement of direct photons in the forward region and 15\% contribution of direct photons in the total neutral meson production validates our claim of additional direct photon contribution. The theoretical predictions from the Golec-Biernat and W$\ddot u$sthoff (GBW) model which predicted the direct photon yield in good agreement with the RHIC data in pp collisions at $\sqrt{\it{s}}$ = 200 GeV, predicts the existence and enhancement of direct photons in the forward region as discussed in the section~\ref{3}. Based on these experimental results and theoretical predictions we argue that the additional contribution of direct photons at $\sqrt{\it{s}_{\rm NN}}$ = 200 GeV in both Cu+Cu and Au+Au collisions are improving the ratio $\frac{N_{ch}}{N_{\gamma}}$ to 1.2 $\pm$ 0.1 from 1.4 $\pm$ 0.1 keeping the energy to charged particle production fraction same.
\section{Acknowledgement}
This work is supported by the Department of Science and Technology (DST), Govt. of India (DST/INSPIRE Fellowship/2018/IF180058). We would like to thank high energy physics group of Department of Physics and Centre for Astroparticle Physics and Space Science, Bose Institute for fruitful discussions.


\begin{thebibliography}{99}
	\bibitem{Ke:2020clc}
	W.~Ke and X.~N.~Wang,``QGP modification to single inclusive jets in a calibrated transport model,'' JHEP \textbf{05}, 041 (2021) \\doi:10.1007/JHEP05(2021)041
	[arXiv:2010.13680 [hep-ph]].
	\bibitem{Goncalves:2019uod}
	K.~J.~Gon\c{c}alves, A.~V.~Giannini, D.~D.~Chinellato and G.~Torrieri,
	``Limiting fragmentation as an initial state probe in heavy ion collisions,''
	Phys. Rev. C \textbf{100}, no.5, 054901 (2019)\\doi:10.1103/PhysRevC.100.054901
	[arXiv:1906.00947 [nucl-th]].
	\bibitem{He:2019vgs}
	M.~He and R.~Rapp, ``Hadronization and Charm-Hadron Ratios in Heavy-Ion Collisions,'' Phys. Rev. Lett. \textbf{124}, no.4, 042301 (2020) \\doi:10.1103/PhysRevLett.124.042301
	[arXiv:1905.09216].
	\bibitem{Cao:2015hia}
	S.~Cao, G.~Y.~Qin and S.~A.~Bass, ``Energy loss, hadronization and hadronic interactions of heavy flavors in relativistic heavy-ion collisions,'' Phys. Rev. C \textbf{92}, no.2, 024907 (2015) \\doi:10.1103/PhysRevC.92.024907 [arXiv:1505.01413 [nucl-th]].
	\bibitem{Casalderrey-Solana:2019ubu}
	J.~Casalderrey-Solana, G.~Milhano, D.~Pablos and K.~Rajagopal, ``Modification of Jet Substructure in Heavy Ion Collisions as a Probe of the Resolution Length of Quark-Gluon Plasma,'' JHEP \textbf{01}, 044 (2020) \\doi:10.1007/JHEP01(2020)044 [arXiv:1907.11248 [hep-ph]].
	\bibitem{PHENIX:2018lia}
	C.~Aidala \textit{et al.} [PHENIX], ``Creation of quark\textendash{}gluon plasma droplets with three distinct geometries,'' Nature Phys. \textbf{15}, no.3, 214-220 (2019) \\doi:10.1038/s41567-018-0360-0 [arXiv:1805.02973 [nucl-ex]].
	\bibitem{Bjorken:1982qr}
	J.~D.~Bjorken, ``Highly Relativistic Nucleus-Nucleus Collisions: The Central Rapidity Region,'' Phys. Rev. D \textbf{27}, 140-151 (1983) \\doi:10.1103/PhysRevD.27.140	 
	\bibitem{Klay:2001tf}
	J.~L.~Klay \textit{et al.} [E895], ``Longitudinal flow from 2-A-GeV to 8-A-GeV Au+Au collisions at the Brookhaven AGS,'' Phys. Rev. Lett. \textbf{88}, 102301 (2002) \\doi:10.1103/PhysRevLett.88.102301 [arXiv:nucl-ex/0111006 [nucl-ex]].
	\bibitem{Gribov:1984tu}
	L.~V.~Gribov, E.~M.~Levin and M.~G.~Ryskin, ``Semihard Processes in QCD,''
	Phys. Rept. \textbf{100}, 1-150 (1983) \\doi:10.1016/0370-1573(83)90022-4
	\bibitem{McLerran:1993ni}
	L.~D.~McLerran and R.~Venugopalan, ``Computing quark and gluon distribution functions for very large nuclei,'' Phys. Rev. D \textbf{49}, 2233-2241 (1994) \\doi:10.1103/PhysRevD.49.2233 [arXiv:hep-ph/9309289 [hep-ph]].
	\bibitem{Abbott:2003ba}
	T.~Abbott, L.~Ahle, Y.~Akiba, D.~Alburger, D.~Beavis, L.~Birstein, M.~A.~Bloomer, P.~D.~Bond, H.~C.~Britt and B.~Budick, \textit{et al.} ``Further observations on midrapidity et distributions with aperture corrected scale,''
	Phys. Rev. C \textbf{68}, 034908 (2003) \\doi:10.1103/PhysRevC.68.034908
	\bibitem{Back:2001bq}
	B.~B.~Back \textit{et al.} [PHOBOS], ``Charged particle pseudorapidity density distributions from Au+Au collisions at $\sqrt{s_{NN}}$ = 130-GeV,'' Phys. Rev. Lett. \textbf{87}, 102303 (2001) \\doi:10.1103/PhysRevLett.87.102303
	[arXiv:nucl-ex/0106006 [nucl-ex]].
	\bibitem{Kharzeev:2001yq}
	D.~Kharzeev, E.~Levin and M.~Nardi, ``The Onset of classical QCD dynamics in relativistic heavy ion collisions,'' Phys. Rev. C \textbf{71}, 054903 (2005) \\doi:10.1103/PhysRevC.71.054903 [arXiv:hep-ph/0111315 [hep-ph]].
	\bibitem{Beckmann:1981gc}
	R.~Beckmann, S.~Raha, N.~Stelte and R.~M.~Weiner, ``Limiting Fragmentation in High-energy Heavy Ion Reactions and Preequilibrium,'' Phys. Lett. B \textbf{105}, 411-416 (1981) \\doi:10.1016/0370-2693(81)91194-1
	\bibitem{Adams:2005cy}
	J.~Adams \textit{et al.} [STAR], ``Multiplicity and pseudorapidity distributions of charged particles and photons at forward pseudorapidity in Au + Au collisions at s(NN)**(1/2) = 62.4-GeV,'' Phys. Rev. C \textbf{73}, 034906 (2006) \\doi:10.1103/PhysRevC.73.034906 [arXiv:nucl-ex/0511026].
	\bibitem{Abelev:2009cy}
	B.~I.~Abelev \textit{et al.} [STAR], ``Center of mass energy and system-size dependence of photon production at forward rapidity at RHIC,'' Nucl. Phys. A \textbf{832}, 134-147 (2010) \\doi:10.1016/j.nuclphysa.2009.11.011 [arXiv:0906.2260 [nucl-ex]].
	\bibitem{Bearden:2003hx}
	I.~G.~Bearden \textit{et al.} [BRAHMS], ``Nuclear stopping in Au + Au collisions at s(NN)**(1/2) = 200-GeV,'' Phys. Rev. Lett. \textbf{93}, 102301 (2004)
	\\doi:10.1103/PhysRevLett.93.102301 [arXiv:nucl-ex/0312023 [nucl-ex]].
	\bibitem{Kharzeev:1996sq}
	D.~Kharzeev, ``Can gluons trace baryon number?,'' Phys. Lett. B \textbf{378}, 238-246 (1996) \\doi:10.1016/0370-2693(96)00435-2 [arXiv:nucl-th/9602027].
	\bibitem{Owens:1986mp}
	J.~F.~Owens, ``Large Momentum Transfer Production of Direct Photons, Jets, and Particles,'' Rev. Mod. Phys. \textbf{59}, 465 (1987) \\doi:10.1103/RevModPhys.59.465
	\bibitem{Linnyk:2013hta}
	O.~Linnyk, V.~P.~Konchakovski, W.~Cassing and E.~L.~Bratkovskaya, ``Photon elliptic flow in relativistic heavy-ion collisions: hadronic versus partonic sources, '' Phys. Rev. C \textbf{88}, 034904 (2013)
	doi:10.1103/PhysRevC.88.034904
	[arXiv:1304.7030 [nucl-th]].
	\bibitem{Albrecht:1995fs}
	R.~Albrecht \textit{et al.} [WA80], ``Limits on the production of direct photons in 200-A/GeV S-32 + Au collisions,'' Phys. Rev. Lett. \textbf{76}, 3506-3509 (1996) \\doi:10.1103/PhysRevLett.76.3506
	\bibitem{Adler:2005ig}
	S.~S.~Adler \textit{et al.} [PHENIX], ``Centrality dependence of direct photon production in s(NN)**(1/ 2) = 200-GeV Au + Au collisions,'' Phys. Rev. Lett. \textbf{94}, 232301 (2005) \\doi:10.1103/PhysRevLett.94.232301 [arXiv:nucl-ex/0503003 [nucl-ex]].
	\bibitem{JalilianMarian:2007bg}
	J.~Jalilian-Marian, ``Rapidity dependence of the photon to pion production ratio in high energy collisions,'' Nucl. Phys. A \textbf{806}, 305-311 (2008)
	\\doi:10.1016/j.nuclphysa.2008.02.300 [arXiv:nucl-th/0703069 [nucl-th]].
	\bibitem{Jeon:2002mf}
	S.~Jeon, J.~Jalilian-Marian and I.~Sarcevic, ``Prompt photon and inclusive pi0 production at RHIC and LHC,'' Nucl. Phys. A \textbf{715}, 795 (2003)
	\\doi:10.1016/S0375-9474(02)01491-4 [arXiv:nucl-th/0211084 [nucl-th]].
	\bibitem{Rezaeian:2009it}
	A.~H.~Rezaeian and A.~Schafer, ``Hadrons and direct photon in pp and pA collisions at LHC and saturation effects,'' Phys. Rev. D \textbf{81}, 114032 (2010) \\doi:10.1103/PhysRevD.81.114032 [arXiv:0908.3695 [hep-ph]].
	\bibitem{Adler:2006yt}
	S.~S.~Adler \textit{et al.} [PHENIX], ``Measurement of direct photon production in p+p collisions at s**(1/2) = 200-GeV,'' Phys. Rev. Lett. \textbf{98}, 012002 (2007) \\doi:10.1103/PhysRevLett.98.012002 [arXiv:hep-ex/0609031].
	\bibitem{Acosta:2004bg}
	D.~Acosta \textit{et al.} [CDF], ``Direct photon cross section with conversions at CDF,'' Phys. Rev. D \textbf{70}, 074008 (2004)
	\\doi:10.1103/PhysRevD.70.074008 [arXiv:hep-ex/0404022].
	\bibitem{Abe:1994rra}
	F.~Abe \textit{et al.} [CDF], ``A Precision measurement of the prompt photon cross-section in $p\bar{p}$ collisions at $\sqrt{s} = 1.8$ TeV,'' Phys. Rev. Lett. \textbf{73}, 2662-2666 (1994) [erratum: Phys. Rev. Lett. \textbf{74}, 1891-1893 (1995)] \\doi:10.1103/PhysRevLett.73.2662.
	\bibitem{Aad:2015lcb}
	G.~Aad \textit{et al.} [ATLAS], ``Centrality, rapidity and transverse momentum dependence of isolated prompt photon production in lead-lead collisions at $\sqrt{s_{\mathrm{NN}}} = 2.76$ TeV measured with the ATLAS detector,''
	Phys. Rev. C \textbf{93}, no.3, 034914 (2016) \\doi:10.1103/PhysRevC.93.034914
	[arXiv:1506.08552].
	\bibitem{Akesson:1989tm}
	T.~\r{A}kesson \textit{et al.} [Helios], ``Inclusive Negative Particle $p(T$) Spectra in $P$ - Nucleus and Nucleus-nucleus Collisions at 200-{GeV} Per Nucleon,'' Z. Phys. C \textbf{46}, 361-368 (1990) \\doi:10.1007/BF01621023
	\bibitem{Albrecht:1990jq}
	R.~Albrecht \textit{et al.} [WA80], ``Upper limit for thermal direct photon production in heavy ion collisions at 60-A/GeV and 200-A/GeV,'' Z. Phys. C \textbf{51}, 1-10 (1991) \\doi:10.1007/BF01579555
\end{thebibliography}
\end{document}